# Renner-Teller Effect on Dissociative Electron Attachments to Carbon Dioxide

Bin Wu, Lei Xia, Yong-Feng Wang, Hong-Kai Li, Xian-Jin Zeng, and Shan Xi Tian*

*Hefei National Laboratory for Physical Sciences at the Microscale and Department of Chemical Physics, University of Science and Technology of China, Hefei, Anhui 230026, China*

* Corresponding author. E-mail: sxtian@ustc.edu.cn

ABSTRACT

Stereo-dynamics of dissociative electron attachments to $CO_2$ is investigated by the $O^-$ anion velocity imaging experiments combined with the *R*-matrix calculations. $^2\Pi_g$ as a Feshbach resonant state of $CO_2^-$ is confirmed to play roles in the dissociations around 8.0 eV. We find that the dynamic evolutions of the Renner-Teller effect lead to the dramatically different anisotropic $O^-$ momentum distributions.

The Renner-Teller (RT) effect arising from the bending vibration couplings with electron in a linear molecule has been studied extensively, providing dynamics information of the potential energy surface (PES) of neutral or cationic species [1,2]. The complex PES, $E_{complex}(R) = E_r(R) - i\Gamma(R)/2$ ($E_r$ is the energy of a resonant state and $\Gamma$ is the resonant width), is another type PES which controls the formation of an electron-molecule resonant state and the subsequent dynamics [3]. However, in the electron-molecule resonant system, also called as the transient negative ion (TNI), very little is known about the dynamic couplings between nucleus and electron motions [4,5]. The RT effect as a typical nucleus-electron coupling is deserved to be investigated, especially on the complex PES of a TNI.

$CO_2$ represents a particularly interesting system for such study because its linear structure ($D_{\infty h}$) at the ground state potentially processes bending when it captures a low-energy electron. It has been found that the vibration excitation was highly selective both via the $^2\Pi_u$ shape resonance $CO_2^-$ around 3.6 eV and via the virtual $CO_2^-$ state formatted in the low-energy electron attachments [5]. The nuclear dynamics associated with the two components $^2A_1$ and $^2B_1$ of the $^2\Pi_u$ resonance due to the RT coupling was further analyzed theoretically [6]. Autodetaching electron and dissociations to anionic and neutral fragments are two predominated decaying channels of TNI. Measurements of the dissociative electron attachments (DEAs) to $CO_2$ have been carried out by the different groups, mainly focusing on the yield efficiency and kinetic energy distributions of $O^-$ [7-13]. The DEA path, $e^- + CO_2(^1\Sigma_g^+) \rightarrow CO_2^-(^2\Pi_u) \rightarrow CO(^1\Sigma^+) + O^-(^2P)$, can be accessed at the low energies (4 – 5 eV) since its thermodynamic threshold lies at 3.99 eV. However, there are long-term arguments about the broad $O^-$ yield peak at 7 – 9eV [7-13]: Is it related to the asymptote of the lowest $CO_2^-$ ($^2\Pi_u$) resonant state and two components $^2A_1$ and $^2B_1$, or another shape resonant state $CO_2^-$ ($^2\Sigma_g^+$) proposed by Claydon et al. [14] and England et al. [15]? However, the latter was disputed and a Feshbach resonant state $^2\Pi_g$ was also suggested [12]. Two dissociation channels to the same products [CO($^1\Sigma^+$) + O$^-$($^2$P)] were postulated via $^2\Sigma_g^+$ at the lower energy while via $^2\Pi_g$ at the higher energy in the Franck-Condon region of the vertical electron attachment [16]. According to the kinetic energy distributions of O$^-$ measured by Chantry [9] and Dressler and Allan [13], some anions exhibit the low kinetic energy (near 0 eV) while the others process ca. 0.6 eV. The sources of these O$^-$ anions with the different kinetic energies are also another debate in the past years, meanwhile it was believed that such distinct dynamic behaviors should be strongly dependent on the complex PES around this energy [17]. Herein, these puzzling misconceptions will be clarified, in particular, the dynamic evolutions of the RT splitting on the complex PES of $CO_2^-$ will be revealed, by the measurements of O$^-$ momentum distribution using our newly developed anion velocity image mapping apparatus [18] and by the *R*-matrix theoretical calculations.

Details of our experimental setup can be found elsewhere [18]. In brief, an effusive molecular beam is perpendicular to the pulsed low-energy electron beam which is emitted from a homemade electron gun; these low-energy electrons are collimated with the homogenous magnetic field (15 – 20 Gauss) produced with a pair of Helmholtz coils (diameter 800 mm). The anion fragments produced in the DEA are periodically (500 Hz) pushed out from the reaction area then fly through the time-of-flight (TOF) tube (installed along the axis of the molecular beam, the total length is 350 mm). Ten electrodes of the TOF mass spectrometer are in charge of the spatial (2×2×2 mm$^3$) and velocity ($\Delta v/v \leq 2.2\%$) focusing of the anions. The anionic fragments produced during one pulse of the electrons will expand in three-dimension (3D) space to form a Newton sphere, and finally they are detected with a pair of micro-channel plates (MCPs) and a phosphor screen. The 3D O$^-$ momentum distributions are directly recorded with a CCD camera using the time-sliced imaging technique [19], namely, a detection time-gate is realized with a high voltage pulse (60 ns width) added on the rear MCP.

As shown in Fig. 1(a), the O⁻ product efficiency curve has been recorded with the electron impact energies less than 10 eV. The spectral profile is exactly same as measured previously [9], thus only two representative points at 4.4 eV and 8.2 eV (red circles) obtained in this work are shown. Here, we record the sliced images of the O⁻ momentum distributions at four typical incident energies, i.e., 4.4, 7.7, 8.2, and 8.7 eV. Obviously, the process $CO_2^-(^2\Pi_u) \rightarrow CO(^1\Sigma^+) + O^-(^2P)$ has been explicitly attributed to the O⁻ production at 4.4 eV [5-13]. Due to the much low kinetic energies of these O⁻ ions (90% less than 0.2 eV) [9], it is difficult to map the anisotropic distribution of the O⁻ momentums which was ever proposed in both the experimental [5] and theoretical studies [6]. The differential cross sections (DCSs) of vibrational excitations around 4 eV indeed show that the DCS at 135° is a little larger than that at 90° [9]. Such anisotropic character is somehow indicated by the ellipse sliced image in Fig. 1b, because the vibration excitations due to the RT splitting at the $CO_2^-(^2\Pi_u)$ resonant state should be closely related to the DEA process. However, the DEA in this low energy range is unsuitable for gaining more insights into the RT effects. We will focus on the DEA dynamics at the second peak of Fig. 1(a).

Firstly, the fixed-nuclei scattering calculations are performed with UK polyatomic R-matrix codes [20]. Gaussian base set 6-311G and the experimental value $R_{C=O}$ = 1.162 Å of this linear molecule [21], but $C_{2v}$ point group are used in the calculations. The position and width of a true resonance can be determined by fitting the eigenphase sum to a Breit-Winger form [22],

$$\delta(E) = \delta_0(E) - \tan^{-1}\left(\frac{\Gamma/2}{E_r - E}\right), \quad (1)$$

where $\delta_0$ is the background phase near the resonance. Two low-lying resonant states $^2\Pi_u$ ($E_r$ = 3.60 eV, $\Gamma$ = 0.4728 eV) and $^2\Pi_g$ ($E_r$ = 8.74 eV, $\Gamma$ = 0.3241 eV) are confirmed by the present $R$-matrix calculations, in which the theoretical $E_r$ values are shifted downward about 0.92 eV for satisfactorily fitting the first shape resonant state $^2\Pi_u$ [5]. The second resonant state $^2\Pi_g$, with a doubly excited configuration $\pi_g^3\sigma^{*2}$ ($\sigma^*$ is Rydberg orbital of $CO_2$), can be identified as a Feshbach resonant state and its parent state is $^3\Pi_g$ of the neutral. No other resonant states are found between these two states. Therefore, this Feshbach state should be responsible for the second O¯ peak observed in Fig. 1(a).

To our surprise, three sliced images recorded at 7.7, 8.2, and 8.7 eV, as shown in Fig. 2(a-c), are distinctly different. A remarkable feature of backward scattering is exhibited at 7.7 eV in Fig. 2(a), although there are another two ion accumulation regions at 60° and 300°. The anisotropic character becomes much clearer with the electron impact energy increases, and the tetrad petal-like pattern appears at 8.2 eV and becomes more distinct at 8.7 eV (see Figs. 2b and 2c). This appealing anisotropic distribution can be interpreted with the parity of the Feshbach resonant state $^2\Pi_g$, as discussed in the following text. It is quite puzzling that what dynamics controls the dramatic changes of the image patterns from 7.7 eV to 8.2 eV. Moreover, at these three energies, we also find the weak O¯ intensities at the center of images. According to the principle of the ion velocity imaging, the kinetic energy distributions of O¯ anions obtained in this work together with the others [9,13,23] for comparison are plotted in Fig. 2(d). A sliced image at 8.1 eV was recorded in a recent parallel study [23], and those experimental data are reproduced and shown as empty circles in Fig. 2(d). The present values are normalized with the O¯ intensities with the kinetic energy near 0 eV. Except for Dressler and Allan's work [13], both of ours and the others [9,23] confirmed the existence of O⁻ ions with the low kinetic energies. These low-energy O⁻ anions may be accompanied with the highly vibrational excited CO ($v$ = 9 - 13) [23].

In the parallel study by Slaughter et al. [23], the sliced image at 8.1 eV is extremely similar to the present image recorded at 7.7 eV (Fig. 2a). The distinct changes of the image pattern in Fig. 2 are reported for the first time. Before revealing the physics underneath the image evolutions, we should recall the basic theories of the angular DCS of anion produced in the DEA process. On the basis of the formulation by O'Malley and Taylor [24], the angular distribution can be determined as,

$$\sigma_{anion}(k,\theta,\phi) \propto \sum_{|\mu|} \left| \sum_{l=|\mu|}^{\alpha} a_{l\mu} Y_{l\mu}(\theta,\phi) \right|^2 \quad (2)$$

where $k$ is the incident electron momentum, $a_{l\mu}(k)$ is the energy dependent expansion coefficient, and $Y_{l\mu}$ is the spherical harmonics. $|\mu| = |\Lambda_f - \Lambda_i|$, representing the difference in the projection of the angular momentum along the internuclear axis for the neutral molecule and the TNI. $l$ is the angular momentum of the incoming electron with values $l \geq |\mu|$. As discussed previously for the $^2\Pi_u$ shape resonant state, two components $^2A_1$ and $^2B_1$ are formed due to the RT splitting [6]. The $^2\Pi_g$ Feshbach resonance may also be influenced by the RT splitting when the vibrational bending mode is excited. Considering the vertical promotion of the electron attachment, in the RT splitting, neither $\Lambda$ nor $l$ is good quantum number, but the vibronic angular momentum $K$ about the axis is good quantum number, $K = |\pm \Lambda + l|$. This is in accord with that obtained by multiplying the electronic and vibrational symmetry species to give the vibronic symmetry species, in this case, for a Π state,

$$\Gamma(\psi_{ev}) = \Pi \times \Pi = \Sigma^+ + \Sigma^- + \Delta, \quad (3)$$

where $\Gamma(\psi_{ev})$ is the irreducible representation of the electron-vibrational wavefunction $\psi_{ev}$ (beyond the Born-Oppenheimer approximation). Σ and Δ states correspond to $K$ = 0 and $K$ = 2, respectively. Assuming the RT effect strongly influences the DEA at 7.7 eV, the angular distribution of O⁻ may be ascribed to the state splitting from the $^2\Pi_g$ Feshbach state to $\Sigma$ and $\Delta$ states, corresponding to $\mu$ = 0 and 2. As shown in Fig. 3(a), we obtain the best fit of the form $\left|aY_{00} + be^{i\delta_1}Y_{10} + ce^{i\delta_2}Y_{20}\right|^2 + \left|dY_{22}\right|^2$ (black line), where $Y_{00}$, $Y_{10}$, and $Y_{20}$ jointly correspond to $\Sigma$ state, while $Y_{22}$ corresponds to $\Delta$ state. The fitted $a : b : c : d$ = 1.0 : 15.2 : 11.6 : 19.1 and the relative phases of the partial waves with respect to that of $s$ ($l$ = 0) wave are $\delta_1$ = 4.485 ($p$ wave) and $\delta_2$ = 1.133 ($d$ wave). This best fitting implies that the $p\Sigma$, $d\Sigma$, and $d\Delta$ scattering amplitudes are predominant at 7.7 eV. The reliability of above fitting can be further proved by the trial with the form $\left|aY_{00} + be^{i\delta_1}Y_{10} + ce^{i\delta_2}Y_{20}\right|^2$ (red line in Fig. 3a) which indicates the serious deviations from the experimental data.

In contrast to the angular momentum distribution of O⁻ at 7.7 eV, the significant anisotropy observed at 8.7 eV cannot be simulated with either of above two forms. As shown in Fig. 3(b), the angular momentum distribution of O⁻ at 8.7 eV can be well fitted using the form $\left|a'Y_{11} + b'e^{i\delta'}Y_{21}\right|^2$ which means a transition from the neutral $^1\Sigma_g^+$ to the anionic resonant state $^2\Pi_g$ ($\mu$ = 1). The fitted values $\delta'$ = 1.947, $a'$ and $b'$ have the ratio of 1: 10.9. This indicates that a $d\Pi$ scattering amplitude is predominant at 8.7 eV. In the similar scenario, the angular momentum distribution of O⁻ at 8.2 eV also basically arises from the $d\Pi$ scatterings although there the forward-backward asymmetry is slightly appearing (see Fig. 2b). It is should be noted that such $d\Pi$ scattering mechanism is completely different from that occurring at 7.7 eV. In the other words, bending vibrations of the molecule are hardly excited during the electron attachment at 8.2 – 8.7 eV. As shown in Fig. 4, the profiles of the PES along the single C=O bond stretching are depicted. Both the $E_r$ value and the width (not shown) of the $^2\Pi_g$ Feshbach state decrease gradually with the elongation of one C=O bond (the TNI structure is still kept as the linear). The electronic ground-state CO and O⁻ can be produced, while the dissociation to CO¯ and O($^1$D) is unfavorable in energy. The present results calculated with the *R*-matrix are in excellent agreement with the recent Kohn scattering calculations [23]. A straightforward interpretation to the forward-backward asymmetry observed at 8.1 eV was also given with the fixed-nuclei complex Kohn scattering calculations, assuming that the bending structure of $CO_2^-$ happened to have the OCO angle of 55° [23].

In summary, the stereo-dynamics of DEA to $CO_2$ is investigated by the O¯ anion velocity imaging experiments combined with the *R*-matrix calculations. We demonstrate the dynamic evolutions of the RT effect by observation of the dramatically different anisotropic O¯ momentum distributions around the Feshbach resonant state $^2\Pi_g$. With the help of the quantum scattering calculations, the anion velocity image mapping technique has been a powerful tool for explorations of the complex PES of TNIs and the related stereo-dynamics.

This work is partially supported by MOST (Grant No. 2011CB921401) and FRFCU (Grant No.WK2340000012).

[24] T. F. O'Malley and H. S. Taylor, Phys. Rev. **176**, 207 (1968).

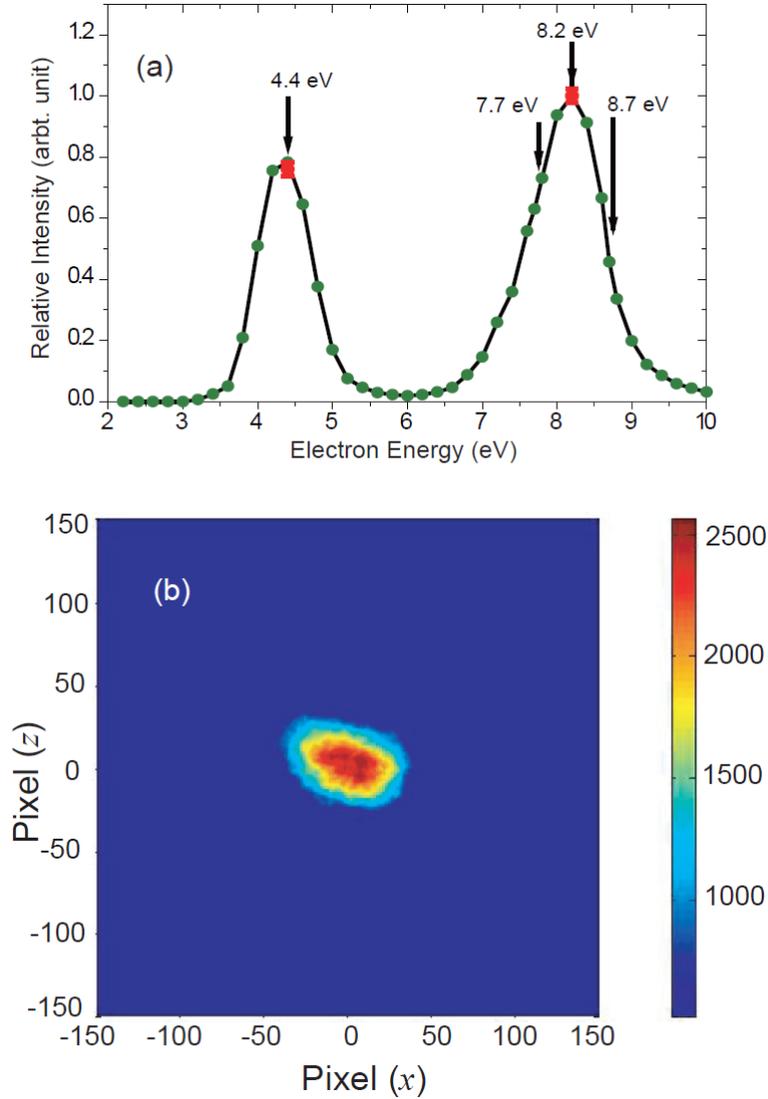

FIG. 1 (color online) (a) The O¯ production efficiency curves: solid circles (green) are adopted from the experimental data at the temperature of 300 K [9]; solid circles with error bars (red) are obtained in this work; the arrows point to the energies at which the sliced images will be recorded. (b) The sliced image of O¯ recorded at the incident energy of 4.4 eV, the electron incident direction is from left to right.

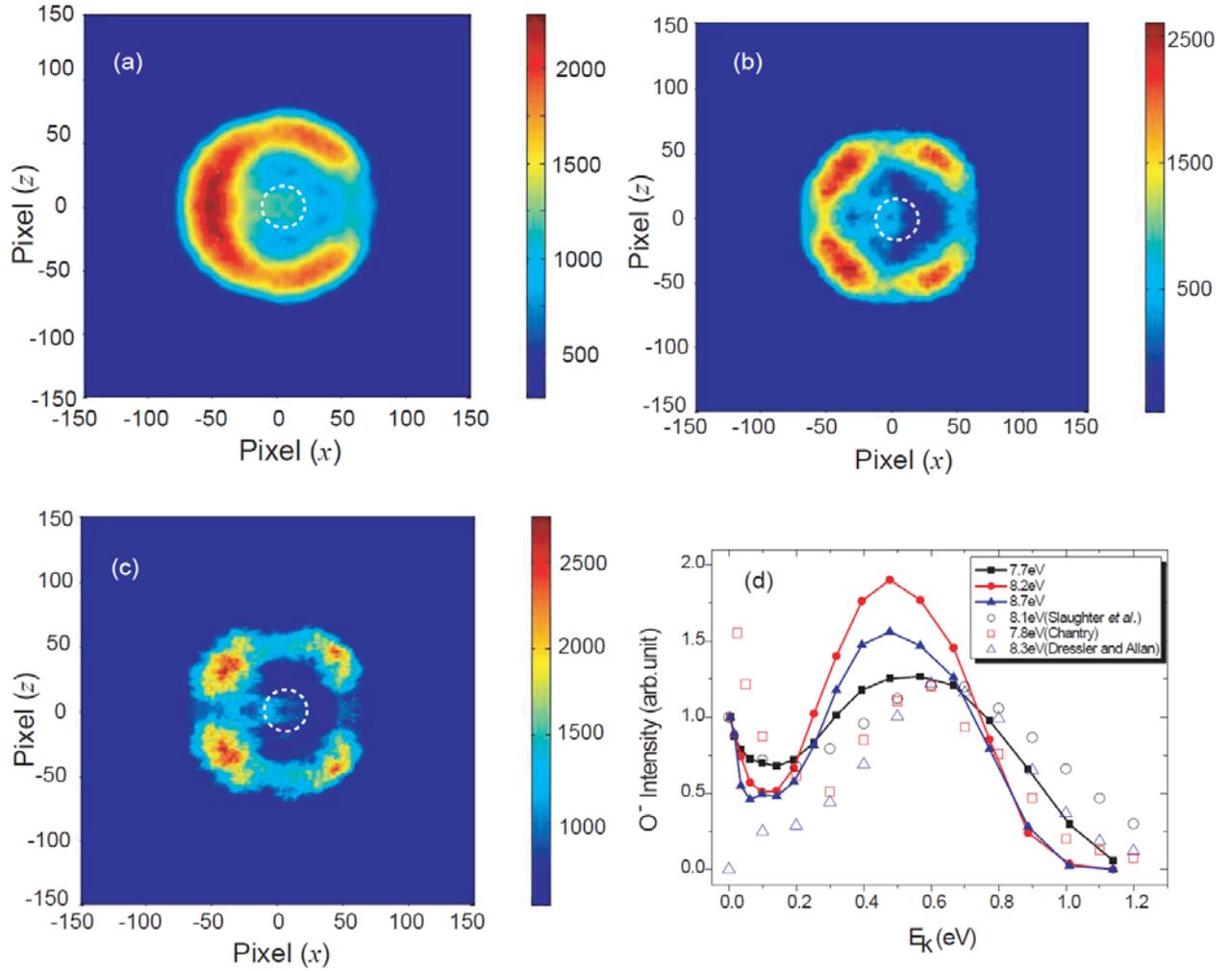

FIG. 2 (color online) (a), (b), and (c) show the sliced images of O¯ recorded at the incident energies of 7.7, 8.2, and 8.7 eV, respectively, in which the small circles (broken white lines) represent the low kinetic-energy ions and the electron incident direction is from left to right. (d) Measured O¯ kinetic energy distributions for the different incident electron energies: The data represented with solid squares (black), solid circles (red), and solid triangles (blue) are obtained in this work. The experimental data of Chantry [9] (empty squares), the uncorrected (empty triangles) data of Dressler and Allan [13], and the recent work (empty circles) of Slaughter *et al*. [23] are adopted for comparison.

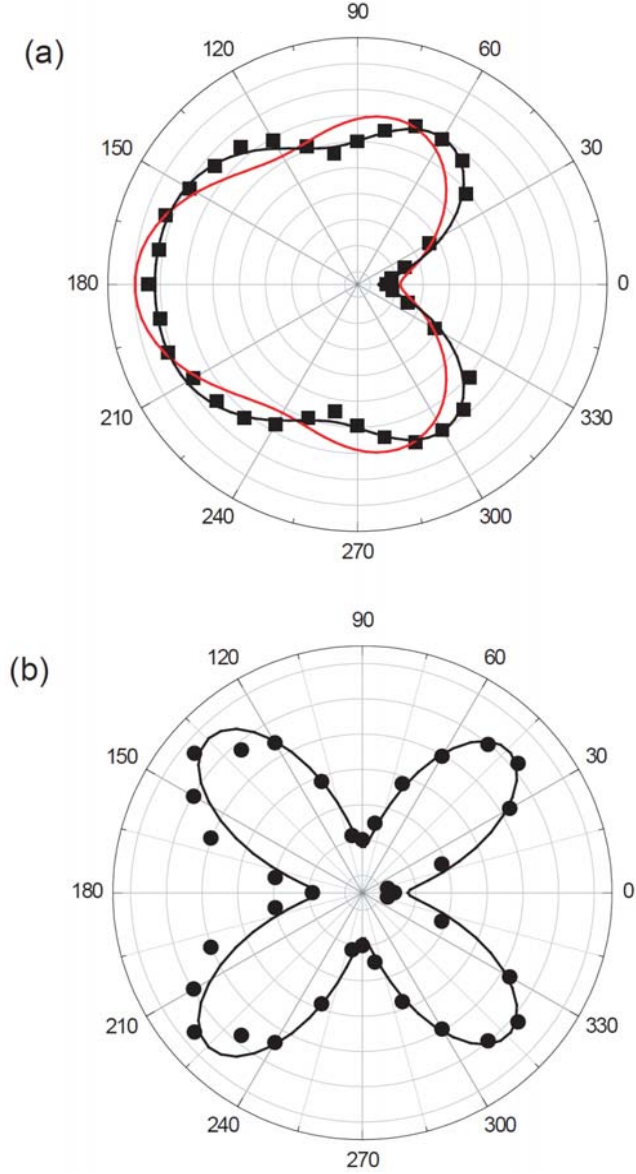

FIG. 3 (color online) (a) Angular distributions of O¯ at the incident energy of 7.7 eV, the black thick-line represents the best fitting with the form $\left| aY_{00} + be^{i\delta_1}Y_{10} + ce^{i\delta_2}Y_{20} \right|^2 + \left| dY_{22} \right|^2$, while the red thick-line represent the fitting with the form $\left| aY_{00} + be^{i\delta_1}Y_{10} + ce^{i\delta_2}Y_{20} \right|^2$. (b) Angular distributions of O¯ at the incident energy of 8.7 eV, the black thick-line represents the best fitting with the form $\left| a'Y_{11} + b'e^{i\delta'}Y_{21} \right|^2$. The experimental data labeled with squares (a) and circles (b) represent the O¯ anions within the kinetic energy range of 0.35- 0.65 eV.

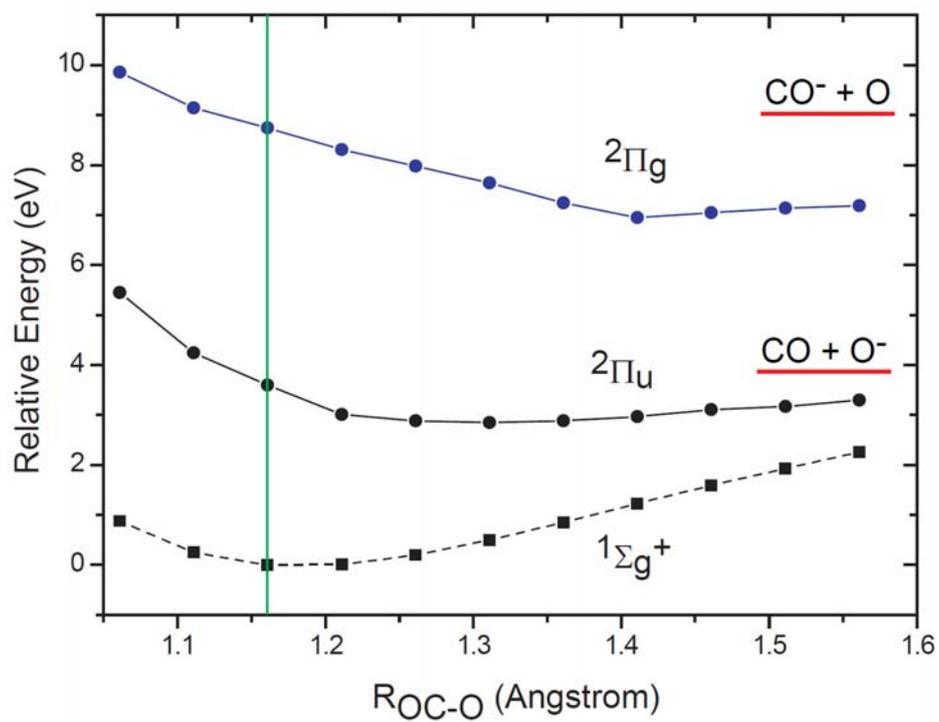

FIG. 4 (color online) Collinear potential energy curves for $CO_2$ (dashed) and $CO_2^-$ (solid). One CO distance is fixed at 1.162 Å. Vertical line (green) indicates equilibrium geometry.